\documentclass[runningheads]{llncs}
\usepackage{graphicx}
\usepackage{amsfonts} 
\usepackage{xcolor}
\usepackage{ltxtable}
\usepackage{multicol}
\usepackage{multirow,url}
\usepackage[misc]{ifsym} 
\usepackage{hyperref}
\usepackage{float,amsmath}
\begin{document}
\title{BiTr-Unet: a CNN-Transformer Combined Network for MRI Brain Tumor Segmentation}
\author{Qiran~Jia \and 
Hai~Shu$^{(\mbox{\scriptsize{\Letter}})}$}
\authorrunning{Q. Jia and H. Shu}
\institute{Department of Biostatistics, School of Global Public Health, New York University, New York, NY 10003, USA \\
\email{hs120@nyu.edu}}
\maketitle              % typeset the header of the contribution
\begin{abstract}
% The abstract should briefly summarize the contents of the paper in 15--250 words.
Convolutional neural networks (CNNs) have achieved remarkable
success in automatically segmenting organs or lesions on 3D medical images. Recently, vision transformer networks have exhibited exceptional performance in 2D image classification tasks. Compared with CNNs, transformer networks have an appealing advantage of extracting long-range features due to their self-attention algorithm. Therefore, we propose a CNN-Transformer combined model, called BiTr-Unet, with specific modifications for brain tumor segmentation on multi-modal MRI scans. 
%In BiTr-Unet, the 3D CNN encoder is responsible for extracting short-range features, and the CNN-decoder reconstructs the segmentation image by collecting residuals from the encoder stages and concatenating them to decrease the loss of features. The fourth and the fifth 3D CNN encoder and decoder stages are connected by a linear projection layer, patch embedding layer to transfer the image tensor to sequence data, ViT layers to extract long-range dependencies, respectively. 
Our BiTr-Unet achieves good performance on the BraTS2021 validation dataset
with median Dice score 0.9335, 0.9304 and 0.8899,
and median Hausdorff distance 2.8284, 2.2361 and 1.4142
for the whole tumor, tumor core, and enhancing tumor, respectively. On the BraTS2021 testing dataset, the corresponding results are 0.9257, 0.9350 and 0.8874 for Dice score, and 3, 2.2361 and 1.4142 for Hausdorff distance. The code is publicly available at \url{https://github.com/JustaTinyDot/BiTr-Unet}.

\keywords{Brain Tumor, Deep Learning, Multi-modal Image Segmentation, Vision Transformer.}
\end{abstract}
\section{Introduction}
As one of the most complicated tasks in computer vision, automated biomedical image segmentation plays an important role in disease diagnosis and further treatment planning. It aims to act like experienced physicians to identify types of tumors and delineate different sub-regions of organs on medical images such as MRI and CT scans \cite{zhong20202wm,lyu2020two}. Earlier segmentation systems were based on traditional approaches such as  edge detection filters, mathematical methods, and machine learning algorithms, but the heavy computational complexity hindered their progress \cite{Hesamian2019Deep}. In recent years, remarkable breakthroughs have been made to computer hardware and deep learning. Deep learning is a derivation of machine learning, which implements multiple processing layers to build a deep neural network to extract representations from data by mimicking the working mechanism of human brains~\cite{Lecun2015DL}. Deep learning has succeeded in various difficult tasks including image classification and speech recognition, and also has become the most prevailing method for automated biomedical image segmentation~\cite{Hesamian2019Deep,NIPS2012_c399862d,Kuniaki2015audio,Shu2021RAI}. While many challenges and limitations such as the shortage of annotated data persist, deep learning exhibits its superiority 
over traditional segmentation methods in processing speed and segmentation accuracy \cite{Hesamian2019Deep}. 

\begin{figure}[t!]
	\centering
	\includegraphics[width=1\textwidth]{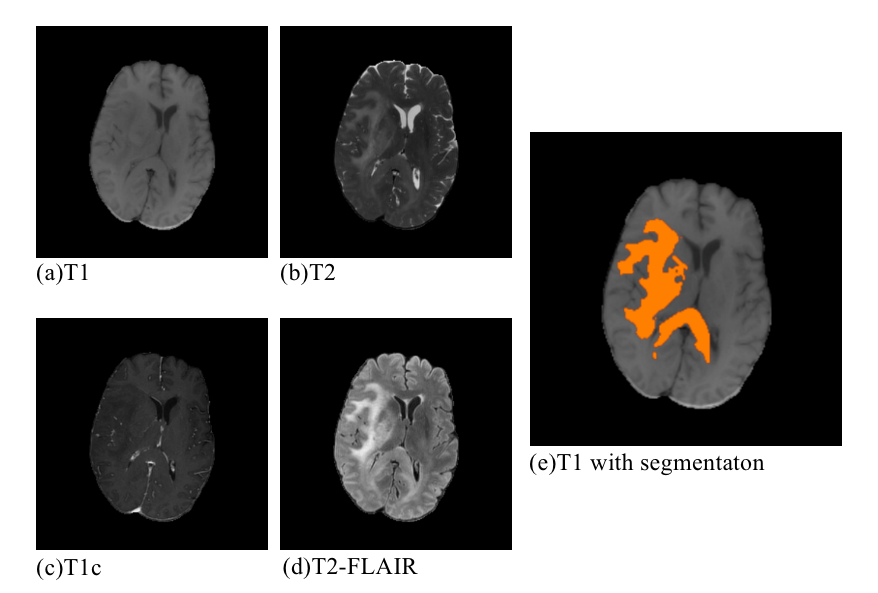}
	\caption{Visualization of one case from the BraTS2021 training data.}
	\label{fig: example}
\end{figure}

To promote the development of biomedical image segmentation techniques, many relevant challenges and conferences are held each year for researchers to propose innovative algorithms and communicate new discoveries. Year 2021 is the tenth anniversary of the Brain Tumor Segmentation Challenge (BraTS), which has been dedicated to being the venue of facilitating the state-of-the-art brain glioma segmentation algorithms~\cite{baid2021rsna,menze2014multimodal,bakas2017advancing,bakas2017segmentation,bakas2017gbm}.Due to privacy protection issues, biomedical image data are notoriously difficult to obtain and usually in a comparatively small scale. However, this year the number of the well-curated multi-institutional multi-parametric MRI scans of glioma provided by the organizer have been updated from 660 to 2000, which largely benefits setting the new benchmark and exploring the potential of algorithms\cite{baid2021rsna}. The training dataset provides four 3D MRI modalities, including native T1-weighted (T1), post-contrast T1-weighted (T1c), T2-weighted (T2), and T2 Fluid Attenuated Inversion Recovery (T2-FLAIR), together with the “ground-truth” tumor segmentation labeled by physicians; see Fig.~\ref{fig: example}. For the validation and testing datasets, the ``ground-truth" tumor segmentation is not open to participants. The ranking criterion in BraTS2021 consists of the Dice score, Hausdorff distance (95\%), Sensitivity, and Specificity for the respective segmentation of the whole tumor, tumor core, and enhancing tumor.

The invention of U-Net, which supplements the contracting network by successive layers to build an encoder-decoder with skip connections architecture, successfully makes the feature extraction in biomedical images more precise and efficient \cite{ronneberger2015u}. U-Net has become the baseline network that is frequently modified for better performance in BraTS ever since, and one of U-Net’s modified versions, nnU-Net, took the first place in BraTS2020 \cite{isensee2018nnu}, which again proves the excellent performance of the fully convolutional network and the structure of successive upsampling layers. Some other top-ranked networks in BraTS2020 also add innovative modules such as the variational autoencoder (VAE), attention gate (AG), as well as Convolutional Block Attention Module (CBAM) into U-net like structures to regularize the feature extraction process and boost generalization performance \cite{isensee2018nnu,henry2020brain,lyu2020two}. Overall, it is generally acknowledged that modifying and enhancing U-Net architecture is an efficient and elegant way to obtain good results for biomedical image segmentation.

Another important network structure, Transformer \cite{vaswani2017attention}, which displays successful improvement on performances of networks in the Natural Language Processing (NLP) realm, is also proved to be useful in the domain of computer vision. Vision Transformer (ViT) \cite{dosovitskiy2020image} proposes the idea that divides an image into several equivalent patches and then applies the self-attention mechanism of Transformer to each patch to capture the long-range dependencies. Compared with networks that are built solely on convolutional layers, ViT overcomes their limitation of locality and therefore makes predictions with more considerations. To boost the performance of ViT and continually set the state-of-the-art benchmark of image classification tasks, various modifications of ViT have been proposed \cite{touvron2021training,chen2021crossvit,wu2021cvt}. Furthermore, to alleviate the high computational cost of the original Transformer on high-resolution images, the architecture design of CNN is introduced to the networks that adopt Transformer \cite{hatamizadeh2021unetr,wang2021transbts}.

Even though many works have been done to apply Transformer to the computer vision realm, only a small number of researches on utilizing Transformer to 3D biomedical image segmentation are available. 
NVIDIA proposed UNETR~\cite{hatamizadeh2021unetr}, which uses a modified ViT as the encoder, and adopts the successive upsampling layers from U-Net as the decoder. Since there are 12 Transformer layers in the encoder of UNETR, the sequence representations of different layers are upsampled by CNN and then concatenated with the decoder layers to create the skip connections. 
Another Transformer-based network designed for 3D biomedical image segmentation is TransBTS \cite{wang2021transbts}, which explores another possibility of incorporating Transformer into CNN-based network. Like a conventional U-Net, there are downsampling and upsampling layers with skip connections in TranBTS, but the unique part is that at the bottom of the network, the encoder and decoder are connected by the following 4 types of layers: a linear projection layer, a patch embedding layer to transfer the image tensor to sequence data, ViT layers to extract long-range dependencies, and a feature mapping layer to fit the sequence data back to the 3D CNN decoder. The benefit of arranging ViT in this way into the U-Net architecture is that at the end of the 3D CNN encoder the ViT can compute long-range dependencies with a global receptive field.

Given that several studies have incorporated Transformer into the task of 3D medical image segmentation and yielded satisfactory results, we believe that such a type of algorithms has the potential to succeed in BraTS2021. TransBTS and UNETR both displayed great potentials to be applied to this challenge with specific modifications and improvement, but TransBTS's arrangement of the ViT layers is more suitable for the BraTS data (we will discuss it in detail in the Method section). Therefore, we decide to borrow the backbone and the core idea from TransBTS to propose a refined version for BraTS2021, called 
 Bi Transformer U-Net (BiTr-Unet). The prefix ``Bi" means that we have two sets of ViT layers in our modified model while TransBTS only has one.

\begin{figure}[b!]
	\centering
	\includegraphics[width=1\textwidth]{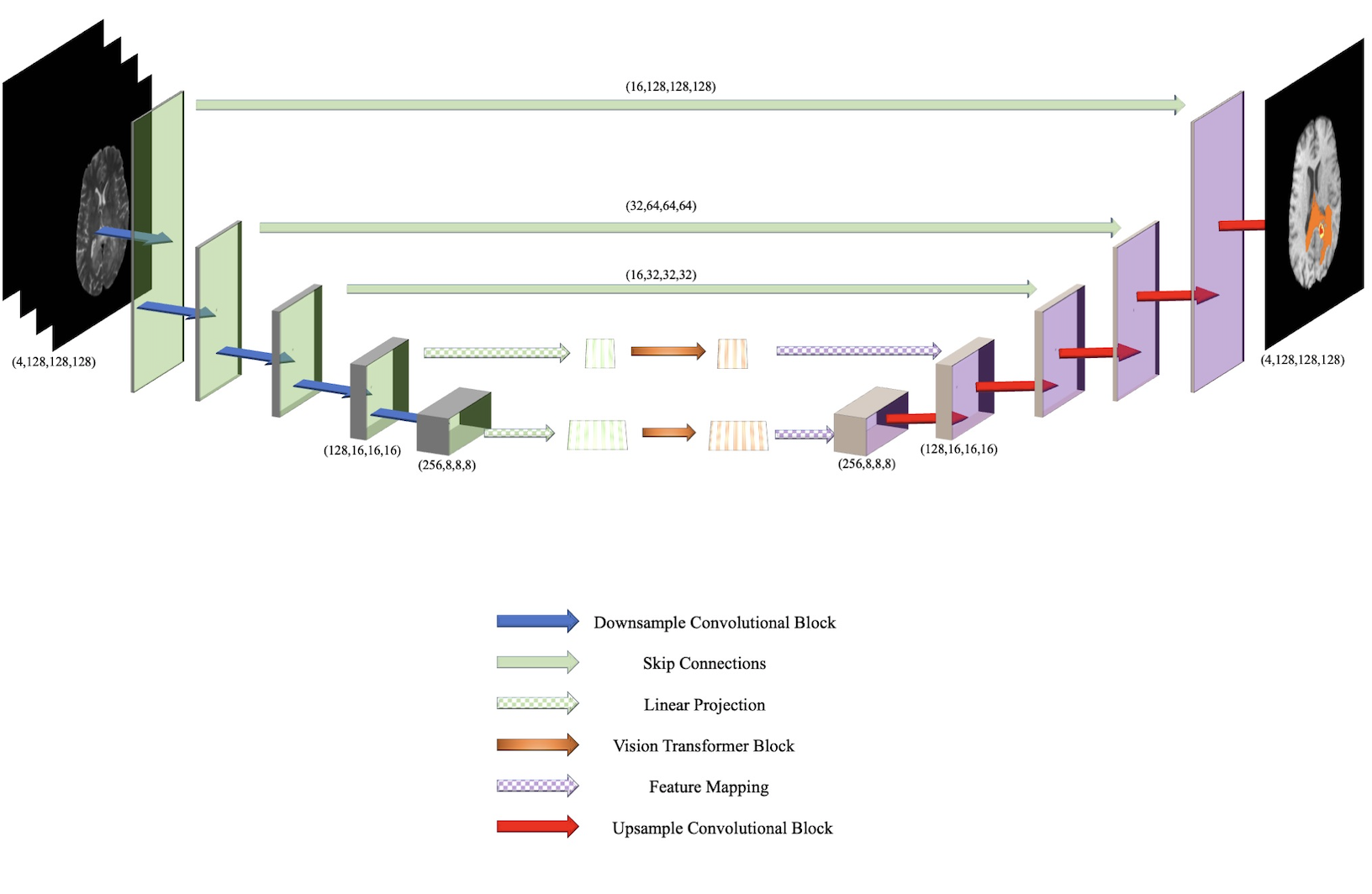}
	\caption{The basic architecture of BiTr-Unet.}
	\label{fig:model_structure}
\end{figure}

\section{Method}\label{Sec: method}
\subsection{Basic Structure of BiTr-Unet}
Fig.~\ref{fig:model_structure} provides an overview of the proposed BiTr-Unet, which contains the main characteristics and core backbone of TransBTS. Overall, the proposed deep neural network $f$ takes the multi-modal MRI scans $X$ as the input, and outputs the predicted segmentation image $Y$, which can be simply written as
\begin{equation}
Y = f(X).
\end{equation}

\subsubsection{Preprocessing}
The BraTS MRI scans with four modalities including T1, T1c, T2, and T2-FLAIR, and the extra ``ground-truth" segmentation label for the training data are stored as separate NIfTI files (.nii.gz). For each patient, the four NIfTI files of MRI images are imported as NumPy arrays and stacked along a new axis to recreate the volumetric information in one NumPy array $X$. For the training data, the NIfTI file of segmentation label is imported as one NumPy array $Y$. Then, $X$, or $X$ and $Y$ for the training data only, will be stored as one Pickle file (.pkl). Details of the implementation of other conventional preprocessing steps are provided in the Implementation Details Section.

\subsubsection{3D CNN Encoder with Attention Module}
BiTr-Unet has an encoder with 3D CNN downsample layers. The preprocessed input of MRIs $X \in \mathbb{R}^{C{\times}H{\times}W{\times}D}$ is fed into the network. After the initial convolutional block with a stride of $1$ to increase the dimension of feature map $F$ to ${4C{\times}H{\times}W{\times}D}$, there are four consecutive  $3{\times}3{\times}3$  convolutional blocks with a stride of $2$ in the encoder to extract the feature representation of the input image. The resulting output dimension of the feature representation is ($64C,\frac{H}{16},\frac{W}{16},\frac{D}{16})$. The Convolutional Block Attention Module (CBAM) is proposed in \cite{woo2018cbam} to improve the performance of CNN models on various tasks. This efficient and lightweight CBAM can be seamlessly integrated into a CNN architecture for adaptive feature refinement by computing attention maps. The original CBAM in \cite{woo2018cbam} is designed for 2D CNN, and in \cite{henry2020brain} the CBAM is expanded for 3D CNN. Given the intermediate feature map $F$, 3D CBAM produces the channel attention map $M_c\in \mathbb{R}^{C{\times}1{\times}1{\times}1{\times}1}$ and the 3D spatial attention map $M_s\in \mathbb{R}^{1{\times}H{\times}W{\times}D}$. The process of computing the 3D spatial attention maps is written as
\begin{align}
F' &= M_c(F) \otimes F, \\
F'' &= M_S(F') \otimes F', 
\end{align}
where $\otimes$ is the element-wise multiplication between maps, and $F''$ represents the refined feature map after the CBAM block. We integrate the 3D CBAM into the encoder by replacing the regular 3D CNN block with 3D CBAM block.

\subsubsection{Feature Embedding of Feature Representation}
Since Transformer takes sequence data as input, the most common way of changing image input into sequence data is to divide an image into several equivalent patches. However, since the input of Transformer in BiTr-Unet is the intermediate feature map $F_{int} \in \mathbb{R}^{K{\times}M{\times}M{\times}M}$, a more convenient way introduced by TransBTS is used. $F_{int}$ firstly go through a $3 {\times} 3 {\times} 3$ convolutional block to increase its channel dimension $K$ to $d$. In this way the spatial and depth dimensions of the feature representations are flattened to one dimension of size $N$. This process is named linear projection, and the resulting feature map $f$ has a dimension of $d {\times} N$, so it can be regarded as $N$ $d$-dimensional tokens to be input of the Transformer block. For the positional embedding of these $N$ $d$-dimensional tokens, TransBTS utilizes learnable position embeddings and adds them to the tokens directly by
\begin{equation}
z_{0}=f +PE=W \times F +PE, \label{embedding}               \end{equation}
where $W$ is the linear projection, $PE\in \mathbb{R}^{d{\times}N}$  is the position embeddings, and $z_{0}\in \mathbb{R}^{d{\times}N}$  is the whole operation of feature embeddings of the feature representations.

\subsubsection{Transformer Layers}
\begin{figure}[b!]
	\centering
	\includegraphics[width=1\textwidth]{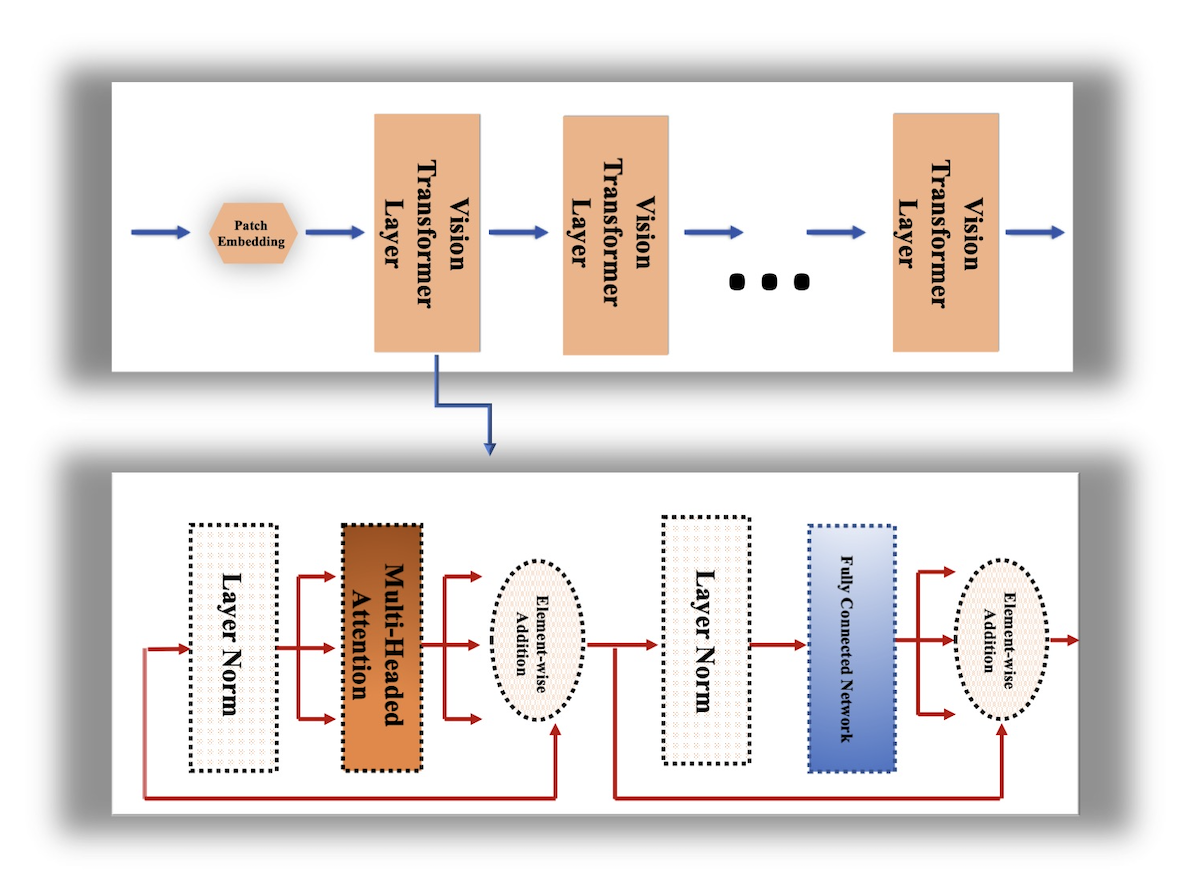}
	\caption{The basic structure of the Vision Transformer Block}
	\label{fig:transformer}
\end{figure}

Fig. \ref{fig:transformer} illustrates the workflow of the Transformer block in our model. After feature embedding, the tokenized and position embedded sequence of feature representations goes through the typical Transformer layers with a Multi-Head Attention (MHA) block and a Feed Forward Network (FFN). The number of Transformer layers $L$ can be modified for the best result. The output of the $l^{th}$ $(l=1,…,L)$ Transformer layer is denoted by $z_l$ and calculated as
\begin{align}
z_l'&=MHA(LN(z_{l-1}))+z_{l-1}, \label{transformer1}  \\
z_l&=FFN(LN(z_{l}' ))+z_l'.  \label{transformer2}  
\end{align}

\subsubsection{Feature Mapping}
Since the outputs of the Transformer layers are the sequence data, it should be transferred back to the intermediate feature representation $F_{int} \in \mathbb{R}^{K{\times}M{\times}M{\times}M}$ for the 3D CNN upsample layer. To achieve this, the output $z_{L}\in \mathbb{R}^{d{\times}N}$ of the Transformer layers are reshaped to $z_{L}\in \mathbb{R}^{d{\times}M{\times}M{\times}M}$ and then go through a $3{\times}3{\times}3$  convolutional block to reduce the channel dimension from $d$ to $K$. After this process, the feature representation $F_{int} \in \mathbb{R}^{K{\times}M{\times}M{\times}M}$  is ready for the 3D CNN upsample layer.

\subsubsection{3D CNN Decoder}
BiTr-Unet has a decoder with 3D CNN upsample layers. There are four consecutive $3{\times}3{\times}3$  convolutional blocks with a stride of 2, and one final convolutional block with a stride of 1 to decrease the dimension of the feature map $F_{final}$ to ${C{\times}H{\times}W{\times}D}$ in the decoder to construct the final segmentation image. The  resulting output segmentation label is $Y\in \mathbb{R}^{C{\times}H{\times}W{\times}D}$.

\subsubsection{Skip Connections}
The output of the first three layers of the 3D CNN layers are directly sent to the last three layers of the 3D CNN layers to create the skip connections. Unlike the output of the first three 3D downsample CNN layers which are concatenated directly with the input of the last three 3D upsample CNN layers, the output of the fourth and the fifth 3D downsample CNN layers go through a feature embedding of feature representation layer, transformer layers, and a feature mapping layer, and then go through the corresponding 3D upsample CNN layer. 

\subsubsection{Postprocessing and Model Ensemble}
The champion model of BraTS2020, nnU-Net, applies a BraTS-specific postprocessing strategy to achieve a higher Dice score, which eliminates a volume of predicted segmentation if this volume is smaller than a threshold \cite{isensee2018nnu}. We borrow this postprocessing technique to maximize the Dice score of the resulting segmentation. Majority voting is an effective and fast method of model ensemble, which may result in a significant improvement in prediction accuracy \cite{lyu2020two}. We adopt majority voting to ensemble differently trained models of BiTr-Unet. Specifically, for each voxel, every model votes for the voxel’s category, and the category with the highest number of votes is used as the final prediction of the voxel. Given $n$ differently trained BiTr-Unet models $f_i(X)$ $(i=1,…,n)$, the output of the majority voting 
for each $j$-th voxel is 
\begin{equation}
C(X)[j]=\text{mode}(f_1(X)[j],...,f_n(X)[j]).
\end{equation}
If the $j$-th voxel has more than one category with the highest number of votes, its final predicted label is the category with the largest averaged 
prediction probability over all trained models, i.e.,
\begin{equation}
y_j= \underset{k\in C(X)[j]}{\arg\max} (p_{jk}), \end{equation}
where $p_{jk} $ is the averaged 
prediction probability of the $k$-th category
for the $j$-th voxel over the $n$ trained models.

\subsection{Comparison with Related Work}
\subsubsection{Modifications from TransBTS}
BiTr-Unet is a variant of TransBTS, but our modifications are significant and make the network well-prepared for BraTS2021. We keep all the innovative and successful attributes of TransBTS, including the combination of CNN and Transformer, feature embedding, and feature mapping. Meanwhile, we also notice that the CNN encoder and decoder in TransBTS can be refined by adding the attention module. We also increase the depth of the whole network for a denser feature representation. TransBTS only utilizes Transformer at the end of the encoder, so it is an operation of the whole feature representation after the fourth layer of the encoder. Since the depth of the encoder layers is increased to five, we use Transformer for both the fourth and the fifth layers. For the fourth layer, Transformer works as an operation to build a skip connection so that its output is concatenated with the input of the fourth decoder layer.

\subsubsection{Differences from UNETR}
A recently proposed network for 3D medical image segmentation, UNETR, also uses Transformer to extract long-range spatial dependencies~\cite{hatamizadeh2021unetr}. Unlike TransBTS and our BiTr-Unet, which involve 3D convolutional blocks in both encoder and decoder of the network, UNETR does not assign any convolutional block to the encoder. ViT \cite{dosovitskiy2020image} is designed to operate on 2D images, and UNETR modifies it 3D images but keeps the original ViT’s design that divides an image into equivalent patches and treats each patch as one token for the operation of attention mechanism. It is an elegant way that using a pure Transformer for the encoding of 3D medical image segmentation, but removing convolutional blocks in the encoder may lead to the insufficiency in extracting local context information for the volumetric BraTS MRI data. Moreover, UNETR stacks Transformer layers and keeps the sequence data dimension unchanged during the whole process, which results in expensive computation for high-resolution 3D images.

\section{Result}

\subsection{Implementation Details}
Our BiTr-Unet model is implemented in Pytorch and trained on four NVIDIA RTX8000 GPUs for 7050 epochs with a batch size of 16. For the optimization, we use the Adam optimizer with an initial learning rate of 0.0002. The learning rate decays by each iteration with a power of 0.9 for better convergence. We adopt this training strategy to avoid overfitting, but it also needs a high number of training epochs. The four modalities of the raw BraTS training data for each case are randomly cropped from the 240 × 240 × 155 voxels to 128×128×128 voxels. We also randomly shift the intensity in the range of [-0.1, 0.1] and scale in the range of [0.9, 1.1]. Test Time Augmentation (TTA),which is proved to increase the prediction accuracy in~\cite{simonyan2015deep}, is applied when using the trained model to generate the segmentation images of the validation data.
During TTA, we create 7 extra copies of the input $X$ by flipping all the possible combinations of directions $H$, $W$ and $D$.

\begin{figure}[t!]
	\centering
	\includegraphics[width=1\textwidth]{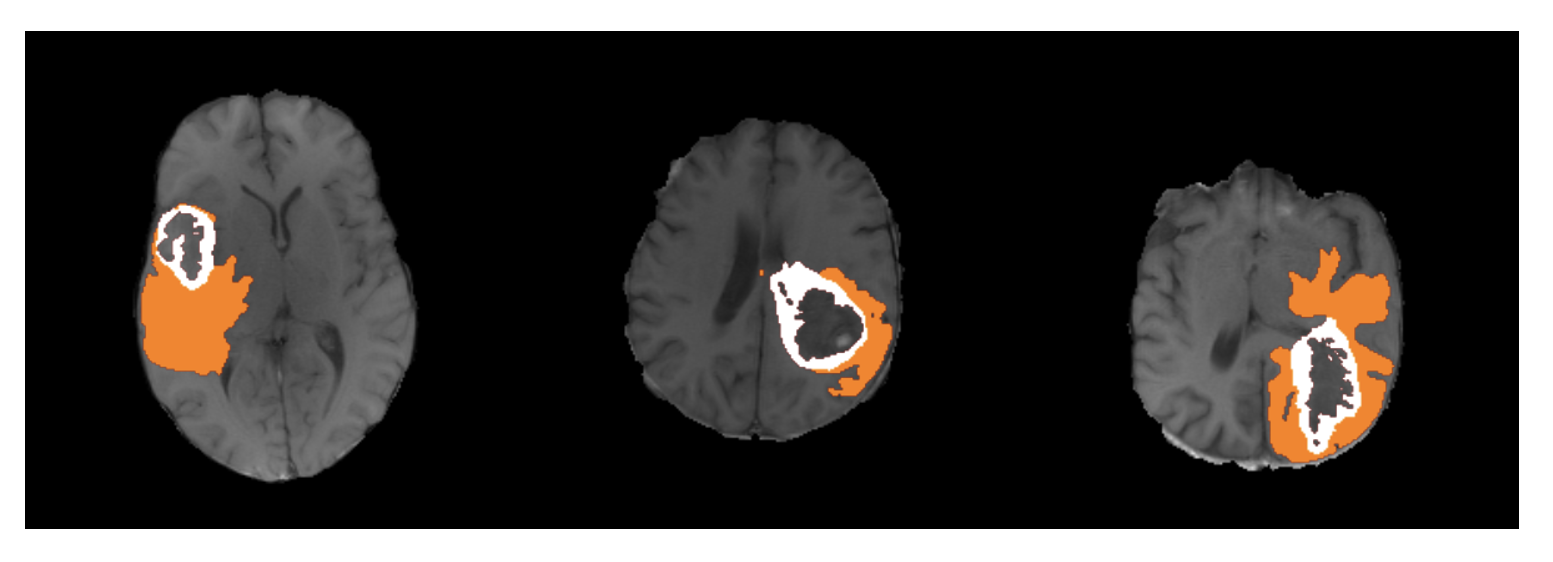}\vspace{-0.8cm}
	\caption{Visualization of the segmentation results from the ensembled BiTr-Unet for three cases in BraTS2021 Validation data.}
	\label{fig: example1}
\end{figure}

\subsection{Segmentation Result}
During the validation phase, the segmentation images of BraTS2021 validation data from our BiTr-Unet are evaluated by the BraTS2021 organizer through the Synapse platform (\url{https://www.synapse.org/#!Synapse:syn25829067/wiki}). We validate three checkpoints with different numbers of trained epochs and also validate the ensemble model of these three checkpoints. Table~\ref{val_result} shows the mean score on the validation data.
Fig.~\ref{fig: example1} presents the segmentation results from the ensembled model for three validation cases.

In the testing phase, we are required to encapsulate the execution environment dependencies and the trained model in an Docker image, and push the built Docker image into Synapse for the testing evaluation. Automated model ensemble through Docker may lead to increased Docker image size,  longer processing time, and even unstable result. To avoid risks, we skip the model ensemble and only select the 7050-epoch BiTr-Unet model for the testing Docker submission. Table~\ref{7050_result} shows the detailed segmentation result of the 7050-epoch BiTr-Unet on BraTS2021 validation data, and Table~\ref{test_result} shows the  result on the testing data.

\begin{table} [H]
	\begin{center}
		\caption{Mean segmentation result of differently trained BiTr-Unet models on BraTS2021 validation data.}\label{val_result}
		
			\begin{tabular}{ c<{\centering}|c|c|c|c|c|c}
			\hline
			&\multicolumn{3}{|c|}{\textbf{Mean Dice}} & \multicolumn{3}{|c}{\textbf{Mean Hausdorff95 (mm)}} \\
			\hline
			\textbf{Model} & \textbf{ET} & \textbf{WT} & \textbf{TC} & \textbf{ET} & \textbf{WT} & \textbf{TC} \\ 
			\hline
			\multirow{1}*{4800-epoch} 
			& 0.8139& 0.9070&0.8466 & 16.7017& 4.5880&13.3968 \\
			\multirow{1}*{6050-epoch} 
			& 0.8237& 0.9079&0.8383 & 15.3014& 4.5883&13.9110 \\
			\multirow{1}*{7050-epoch} 
			& 0.8187& 0.9097&0.8434 & 17.8466& 4.5084&16.6893 \\
			\multirow{1}*{Ensemble by Majority Voting} 
			& 0.8231& 0.9076&0.8392 & 14.9963& 4.5322&13.4592 \\
			\hline 
		\end{tabular}
	\end{center}
\end{table}

\begin{table} [H]
	\begin{center}
		\caption{Detailed segmentation result of the 7050-epoch BiTr-Unet model on BraTS2021 validation data.}\label{7050_result}
		
			\begin{tabular}{ c<{\centering}|c|c|c|c|c|c}
			\hline
			&\multicolumn{3}{|c|}{\textbf{Dice}} & \multicolumn{3}{|c}{\textbf{Hausdorff95 (mm)}} \\
			\hline
			\textbf{Statistics} & \textbf{ET} & \textbf{WT} & \textbf{TC} & \textbf{ET} & \textbf{WT} & \textbf{TC} \\ \hline
			\multirow{1}*{Mean} 
			& 0.8187& 0.9097&0.8434 & 17.8466& 4.5084&16.6893 \\
			\multirow{1}*{Standard deviation} 
			& 0.2327& 0.08837&0.2305 & 73.6517& 7.6621&65.5411 \\
			\multirow{1}*{Median} 
			& 0.8899& 0.9366&0.9338 & 1.4142& 2.8284&2.2361 \\
			\multirow{1}*{25th percentile} 
			& 0.8289& 0.8903&0.8610 & 1& 1.4142&1 \\
			\multirow{1}*{75th percentile} 
			& 0.9397& 0.9584&0.9616 & 2.6390& 4.4721&4.5826 \\
			\hline 
		\end{tabular}
	\end{center}\end{table}

\begin{table} [H]
	\begin{center}
		\caption{Detailed segmentation result of the 7050-epoch BiTr-Unet model on BraTS2021 testing data.}\label{test_result}
		
			\begin{tabular}{ c<{\centering}|c|c|c|c|c|c}
			\hline
			&\multicolumn{3}{|c|}{\textbf{Dice}} & \multicolumn{3}{|c}{\textbf{Hausdorff95 (mm)}} \\
			\hline
			\textbf{Statistics} & \textbf{ET} & \textbf{WT} & \textbf{TC} & \textbf{ET} & \textbf{WT} & \textbf{TC} \\ \hline
			\multirow{1}*{Mean} 
			& 0.7256& 0.7639&0.7422 & 65.2966& 62.1576&69.006318 \\
			\multirow{1}*{Standard deviation} 
			& 0.3522& 0.3426&0.3738 & 139.003958& 133.4915&140.8401 \\
			\multirow{1}*{Median} 
			& 0.8874& 0.9257&0.9350 & 1.4142& 3&2.2361 \\
			\multirow{1}*{25th percentile} 
			& 0.7313& 0.8117&0.7642 & 1& 1.4142&1 \\
			\multirow{1}*{75th percentile} 
			& 0.9512& 0.9600&0.9708 & 3.7417& 9.1647&8.2002 \\
			\hline 
		\end{tabular}
	\end{center}
\end{table}

\section{Discussion}
We propose a new deep neural network model called BiTr-Unet, a refined version of TransBTS, for the BraTS2021 tumor segmentation challenge. The result on the validation data indicates that BiTr-Unet is a stable and powerful network to extract both local and long-range dependencies of 3D MRI scans. Compared to models with U-Net alike architectures but without attention modules that are successful in BraTS2020, BiTr-Unet takes the advantage of the novel Transformer module for potentially better performance. The way we incorporate Transformer into a CNN encoder-decoder model is inspired by TransBTS, but we propose more innovative and suitable modifications such as making feature representations of the skip connection to go through a ViT block. However, there is plenty more room for further explorations to incorporate Transformer into networks for 3D medical image segmentation. The result on the testing data shares similar median and 75th percentile with that on the validation data, but has an inferior mean,standard deviation, and 25th percentile. With the fact that testing data is more dissimilar to the training data than the validation data, this discrepancy result indicates that BiTr-Unet may still struggle when processing unseen patterns. Due to the time limit, we did not finish the experiment to test the performance of the model with more encoder and decoder layers that are connected by a skip connection going through a ViT Block, or to test ViT with different embedding dimensions. Also, inspired by the recent work of innovative Transformer models for image classification or segmentation that reduces computation complexity by special algorithms \cite{xie2021cotr,lin2021cat,liu2021swin} , we would also try to introduce them into our model in the future to improve the performance and decrease the computation complexity.

\section{Acknowledgements}
This research was partially supported by the grant R21AG070303 from the National Institutes of Health and a startup fund from New York University. The
content is solely the responsibility of the authors and does not necessarily represent the official views of the National Institutes of Health or New York University.

\bibliographystyle{splncs04.bst}
\bibliography{report}

\begin{thebibliography}{10}
\providecommand{\url}[1]{\texttt{#1}}
\providecommand{\urlprefix}{URL }
\providecommand{\doi}[1]{https://doi.org/#1}

\bibitem{baid2021rsna}
Baid, U., Ghodasara, S., Bilello, M., Mohan, S., Calabrese, E., Colak, E.,
  Farahani, K., Kalpathy-Cramer, J., Kitamura, F.C., Pati, S., et~al.: The
  rsna-asnr-miccai brats 2021 benchmark on brain tumor segmentation and
  radiogenomic classification. arXiv preprint arXiv:2107.02314  (2021)

\bibitem{bakas2017gbm}
Bakas, S., Akbari, H., Sotiras, A., Bilello, M., Rozycki, M., Kirby, J.,
  Freymann, J., Farahani, K., Davatzikos, C.: Segmentation labels and radiomic
  features for the pre-operative scans of the tcga-gbm collection (07 2017).
  \doi{10.7937/K9/TCIA.2017.KLXWJJ1Q}

\bibitem{bakas2017segmentation}
Bakas, S., Akbari, H., Sotiras, A., Bilello, M., Rozycki, M., Kirby, J.,
  Freymann, J., Farahani, K., Davatzikos, C.: Segmentation labels and radiomic
  features for the pre-operative scans of the tcga-lgg collection. The cancer
  imaging archive  \textbf{286} (2017)

\bibitem{bakas2017advancing}
Bakas, S., Akbari, H., Sotiras, A., Bilello, M., Rozycki, M., Kirby, J.S.,
  Freymann, J.B., Farahani, K., Davatzikos, C.: Advancing the cancer genome
  atlas glioma mri collections with expert segmentation labels and radiomic
  features. Scientific data  \textbf{4}(1),  1--13 (2017)

\bibitem{chen2021crossvit}
Chen, C.F., Fan, Q., Panda, R.: Crossvit: Cross-attention multi-scale vision
  transformer for image classification. arXiv preprint arXiv:2103.14899  (2021)

\bibitem{dosovitskiy2020image}
Dosovitskiy, A., Beyer, L., Kolesnikov, A., Weissenborn, D., Zhai, X.,
  Unterthiner, T., Dehghani, M., Minderer, M., Heigold, G., Gelly, S., et~al.:
  An image is worth 16x16 words: Transformers for image recognition at scale.
  arXiv preprint arXiv:2010.11929  (2020)

\bibitem{hatamizadeh2021unetr}
Hatamizadeh, A., Yang, D., Roth, H., Xu, D.: Unetr: Transformers for 3d medical
  image segmentation. arXiv preprint arXiv:2103.10504  (2021)

\bibitem{henry2020brain}
Henry, T., Carre, A., Lerousseau, M., Estienne, T., Robert, C., Paragios, N.,
  Deutsch, E.: Brain tumor segmentation with self-ensembled, deeply-supervised
  3d u-net neural networks: a brats 2020 challenge solution. arXiv preprint
  arXiv:2011.01045  (2020)

\bibitem{Hesamian2019Deep}
Hesamian, M.H., Jia, W., He, X., Kennedy, P.: Deep learning techniques for
  medical image segmentation: Achievements and challenges. Journal of Digital
  Imaging  \textbf{32} (05 2019). \doi{10.1007/s10278-019-00227-x}

\bibitem{isensee2018nnu}
Isensee, F., Petersen, J., Klein, A., Zimmerer, D., Jaeger, P.F., Kohl, S.,
  Wasserthal, J., Koehler, G., Norajitra, T., Wirkert, S., et~al.: nnu-net:
  Self-adapting framework for u-net-based medical image segmentation. arXiv
  preprint arXiv:1809.10486  (2018)

\bibitem{NIPS2012_c399862d}
Krizhevsky, A., Sutskever, I., Hinton, G.E.: Imagenet classification with deep
  convolutional neural networks. In: Pereira, F., Burges, C.J.C., Bottou, L.,
  Weinberger, K.Q. (eds.) Advances in Neural Information Processing Systems.
  vol.~25. Curran Associates, Inc. (2012),
  \url{https://proceedings.neurips.cc/paper/2012/file/c399862d3b9d6b76c8436e924a68c45b-Paper.pdf}

\bibitem{Lecun2015DL}
LeCun, Y., Bengio, Y., Hinton, G.: Deep learning. Nature  \textbf{521},
  436--44 (05 2015). \doi{10.1038/nature14539}

\bibitem{lin2021cat}
Lin, H., Cheng, X., Wu, X., Yang, F., Shen, D., Wang, Z., Song, Q., Yuan, W.:
  Cat: Cross attention in vision transformer. arXiv preprint arXiv:2106.05786
  (2021)

\bibitem{liu2021swin}
Liu, Z., Lin, Y., Cao, Y., Hu, H., Wei, Y., Zhang, Z., Lin, S., Guo, B.: Swin
  transformer: Hierarchical vision transformer using shifted windows. arXiv
  preprint arXiv:2103.14030  (2021)

\bibitem{lyu2020two}
Lyu, C., Shu, H.: A two-stage cascade model with variational autoencoders and
  attention gates for mri brain tumor segmentation. In: Crimi, A., Bakas, S.
  (eds.) Brainlesion: Glioma, Multiple Sclerosis, Stroke and Traumatic Brain
  Injuries. pp. 435--447. Springer International Publishing, Cham (2021)

\bibitem{menze2014multimodal}
Menze, B.H., Jakab, A., Bauer, S., Kalpathy-Cramer, J., Farahani, K., Kirby,
  J., Burren, Y., Porz, N., Slotboom, J., Wiest, R., et~al.: The multimodal
  brain tumor image segmentation benchmark (brats). IEEE transactions on
  medical imaging  \textbf{34}(10),  1993--2024 (2014)

\bibitem{Kuniaki2015audio}
Noda, K., Yamaguchi, Y., Nakadai, K., Okuno, H., Ogata, T.: Audio-visual speech
  recognition using deep learning. Applied Intelligence  \textbf{42}(4),
  722--737 (Jun 2015). \doi{10.1007/s10489-014-0629-7}

\bibitem{ronneberger2015u}
Ronneberger, O., Fischer, P., Brox, T.: U-net: Convolutional networks for
  biomedical image segmentation. In: International Conference on Medical image
  computing and computer-assisted intervention. pp. 234--241. Springer (2015)

\bibitem{Shu2021RAI}
Shu, H., Chiang, T., Wei, P., Do, K.A., Lesslie, M.D., Cohen, E.O., Srinivasan,
  A., Moseley, T.W., Chang~Sen, L.Q., Leung, J.W.T., Dennison, J.B., Hanash,
  S.M., Weaver, O.O.: A deep learning approach to re-create raw full-field
  digital mammograms for breast density and texture analysis. Radiology:
  Artificial Intelligence  \textbf{3}(4),  e200097 (2021).
  \doi{10.1148/ryai.2021200097}

\bibitem{simonyan2015deep}
Simonyan, K., Zisserman, A.: Very deep convolutional networks for large-scale
  image recognition (2015)

\bibitem{touvron2021training}
Touvron, H., Cord, M., Douze, M., Massa, F., Sablayrolles, A., J{\'e}gou, H.:
  Training data-efficient image transformers \& distillation through attention.
  In: International Conference on Machine Learning. pp. 10347--10357. PMLR
  (2021)

\bibitem{vaswani2017attention}
Vaswani, A., Shazeer, N., Parmar, N., Uszkoreit, J., Jones, L., Gomez, A.N.,
  Kaiser, {\L}., Polosukhin, I.: Attention is all you need. In: Advances in
  neural information processing systems. pp. 5998--6008 (2017)

\bibitem{wang2021transbts}
Wang, W., Chen, C., Ding, M., Li, J., Yu, H., Zha, S.: Transbts: Multimodal
  brain tumor segmentation using transformer. arXiv preprint arXiv:2103.04430
  (2021)

\bibitem{woo2018cbam}
Woo, S., Park, J., Lee, J.Y., Kweon, I.S.: Cbam: Convolutional block attention
  module. In: Proceedings of the European conference on computer vision (ECCV).
  pp. 3--19 (2018)

\bibitem{wu2021cvt}
Wu, H., Xiao, B., Codella, N., Liu, M., Dai, X., Yuan, L., Zhang, L.: Cvt:
  Introducing convolutions to vision transformers. arXiv preprint
  arXiv:2103.15808  (2021)

\bibitem{xie2021cotr}
Xie, Y., Zhang, J., Shen, C., Xia, Y.: Cotr: Efficiently bridging cnn and
  transformer for 3d medical image segmentation. arXiv preprint
  arXiv:2103.03024  (2021)

\bibitem{zhong20202wm}
Zhong, L., Li, T., Shu, H., Huang, C., Johnson, J.M., Schomer, D.F., Liu, H.L.,
  Feng, Q., Yang, W., Zhu, H.: 2wm: Tumor segmentation and tract statistics for
  assessing white matter integrity with applications to glioblastoma patients.
  NeuroImage  \textbf{223},  117368 (2020)

\end{thebibliography}

\end{document}